\documentclass[journal]{IEEEtran}
\usepackage{amsmath,amsfonts}
\usepackage{algorithmic}
\usepackage{algorithm}
\usepackage{array}
\usepackage[caption=false,font=normalsize,labelfont=sf,textfont=sf]{subfig}
\usepackage{textcomp}
\usepackage{stfloats}
\usepackage{url}
\usepackage{verbatim}
\usepackage{graphicx}
\usepackage{cite}
\usepackage{graphicx}
\usepackage{amsmath}
\usepackage{amssymb}
\usepackage{booktabs}

\usepackage{algorithm}
\usepackage{algorithmic}
\usepackage[normalem]{ulem}
\useunder{\uline}{\ul}{}
\usepackage{booktabs}

\usepackage{multirow}
\usepackage{booktabs} 
\usepackage[dvipsnames, svgnames, x11names]{xcolor} 
\usepackage{color}
\usepackage{colortbl}  %color table
\newcommand{\etal} {\textit{et al.}}

\newcommand{\green}[1]{{\textcolor{Green}{ #1}}}
\newcommand{\red}[1]{{\textcolor{Red}{ #1}}}
\begin{document}

\title{MaskBlur: Spatial and Angular Data Augmentation for Light Field Image Super-Resolution}
% \title{MaskBlur: Boosting Spatial and Angular Domain for Light Field Image Super-Resolution}

% \author{Wentao Chao,~\IEEEmembership{Staff,~IEEE,}
\author{Wentao Chao, Fuqing Duan, Yulan Guo,~\IEEEmembership{Senior Member,~IEEE}, Guanghui Wang,~\IEEEmembership{Senior Member,~IEEE}
        % <-this % stops a space
\thanks{W. Chao and F. Duan are with the School of
Artificial Intelligence, Beijing Normal University, Beijing 100875, China. (e-mail: chaowentao@mail.bnu.edu.cn; fqduan@bnu.edu.cn).}% <-this % stops a space
% \thanks{Y. Wang is with the College of Electronic Science and Technology, National University of Defense Technology, Changsha 410073, China. (e-mail: wangyingqian16@nudt.edu.cn).}% <-this % stops a space
\thanks{Y. Guo is with the School of Electronics and Communication
Engineering, Sun Yat-sen University, Guangzhou 510275, China. (e-mail:
guoyulan@sysu.edu.cn)}
\thanks{G. Wang is with the Department of Computer Science, Toronto Metropolitan University, Toronto, ON M5B 2K3, Canada. (e-mail: wangcs@torontomu.ca).}
\thanks{This work is supported by the National Key Research and Development Project Grant, Grant/Award Number: 2018AAA0100802.}
% % Toronto Metropolitan University University of Kansas
% % \thanks{This paper was recommended by Associate Editor xx. (Corresponding author: xx.)}
% \thanks{Corresponding author: F. Duan}
% \thanks{Manuscript received February 3, 2023; revised August 16, 2021.}}
\thanks{Manuscript received May 31, 2024; revised August 13, 2024; accepted September 06, 2024. Corresponding author: F. Duan}
}

% {Shell \MakeLowercase{\etal}: A Sample Article Using IEEEtran.cls for IEEE Journals}


\maketitle

\begin{abstract}
    Data augmentation (DA) is an effective approach for enhancing model performance with limited data, such as light field (LF) image super-resolution (SR). LF images inherently possess rich spatial and angular information. Nonetheless, there is a scarcity of DA methodologies explicitly tailored for LF images, and existing works tend to concentrate solely on either the spatial or angular domain.
    This paper proposes a novel spatial and angular DA strategy named MaskBlur for LF image SR by concurrently addressing spatial and angular aspects. MaskBlur consists of spatial blur and angular dropout two components. Spatial blur is governed by a spatial mask, which controls where pixels are blurred, i.e., pasting pixels between the low-resolution and high-resolution domains. The angular mask is responsible for angular dropout, i.e., selecting which views to perform the spatial blur operation. By doing so, MaskBlur enables the model to treat pixels differently in the spatial and angular domains when super-resolving LF images rather than blindly treating all pixels equally. Extensive experiments demonstrate the efficacy of MaskBlur in significantly enhancing the performance of existing SR methods. We further extend MaskBlur to other LF image tasks such as denoising, deblurring, low-light enhancement, and real-world SR. Code is publicly available at \url{https://github.com/chaowentao/MaskBlur}.
\end{abstract}


\section{Introduction}
\label{sec:intro}

\IEEEPARstart{L}{ight} Field (LF) \cite{ghassab2019light, liu2021intra, sun2022learning, liu2022efficient, cong2023exploiting} images capture spatial and angular information of a scene in a single shot using an LF camera, giving rise to various important applications such as refocusing, depth estimation~\cite{wang2022occlusion, sheng2023lfnat, chao2023learning}, view synthesis \cite{wu2021spatial}, and 3D reconstruction \cite{kim2013scene}. 
However, due to the inherent limitations of commercialized LF cameras, there is a trade-off between spatial and angular resolution in LF images. To enhance the spatial resolution of LF images, traditional methods typically rely on the geometric information of the LF and manually designed features for super-resolution (SR), which have limited performance and slow processing speed. In recent years, deep learning-based LF image SR methods~\cite{liang2022light, wang2022disentangling, cheng2022spatial, chen2022light, van2023light, wang2023ntire, wang2024real, NTIRE2024-LFSR}
%~\cite{bishop2011light,mitra2012light,wanner2013variational,farrugia2017super,alain2018light,rossi2018geometry,yoon2017light, jin2020light, zhang2021end, zhang2019residual, wang2018lfnet, liang2022light, wang2022disentangling, wang2020spatial, yeung2018light, meng2019high, meng2020high, cheng2021light, cheng2022spatial, chen2022light, van2023light, wang2023ntire, wang2024real, NTIRE2024-LFSR}
have made significant progress in terms of accuracy and speed. Nevertheless, the performance of these methods is still constrained by the available dataset size, and collecting a large-scale LF dataset is time-consuming and labor-intensive. Therefore, exploring efficient data augmentation (DA) techniques specifically for LF images SR is highly desirable.

\begin{figure}[tb]
  \centering
  \includegraphics[width=\linewidth]{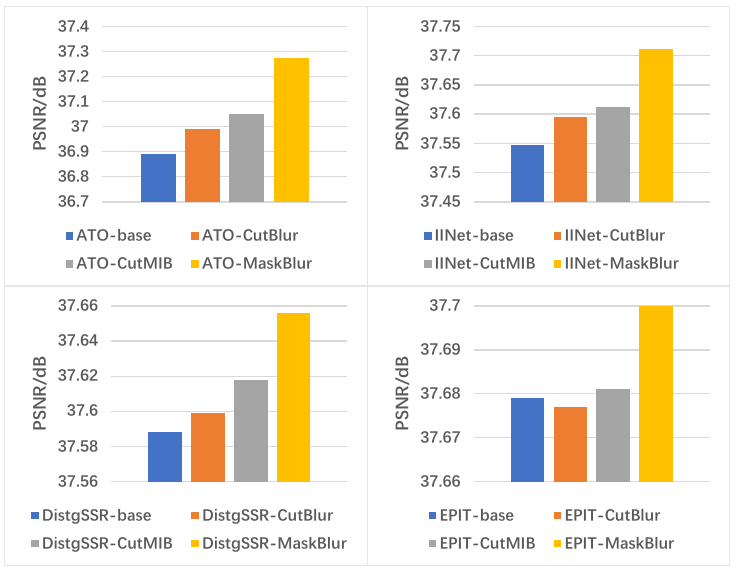}
  \caption{Comparison MaskBlur with CutBlur~\cite{yoo2020rethinking} and CutMIB~\cite{xiao2023cutmib} on various LF image SR methods, i.e., ATO~\cite{jin2020light}, , IINet~\cite{liu2021intra}, DistgSSR ~\cite{wang2022disentangling} and EPIT~\cite{liang2023learning}. We compare the average PSNR (dB, $\uparrow$) on the HCIold~\cite{wanner2013datasets} dataset. Note that MaskBlur improves the values of PSNR by a large margin compared to other DA schemes on various LF image SR methods.}
  \label{fig:intro_chart}
\end{figure}

% InterNet \cite{wang2020spatial}

\begin{figure*}[tb]
  \centering
  \includegraphics[width=\linewidth]{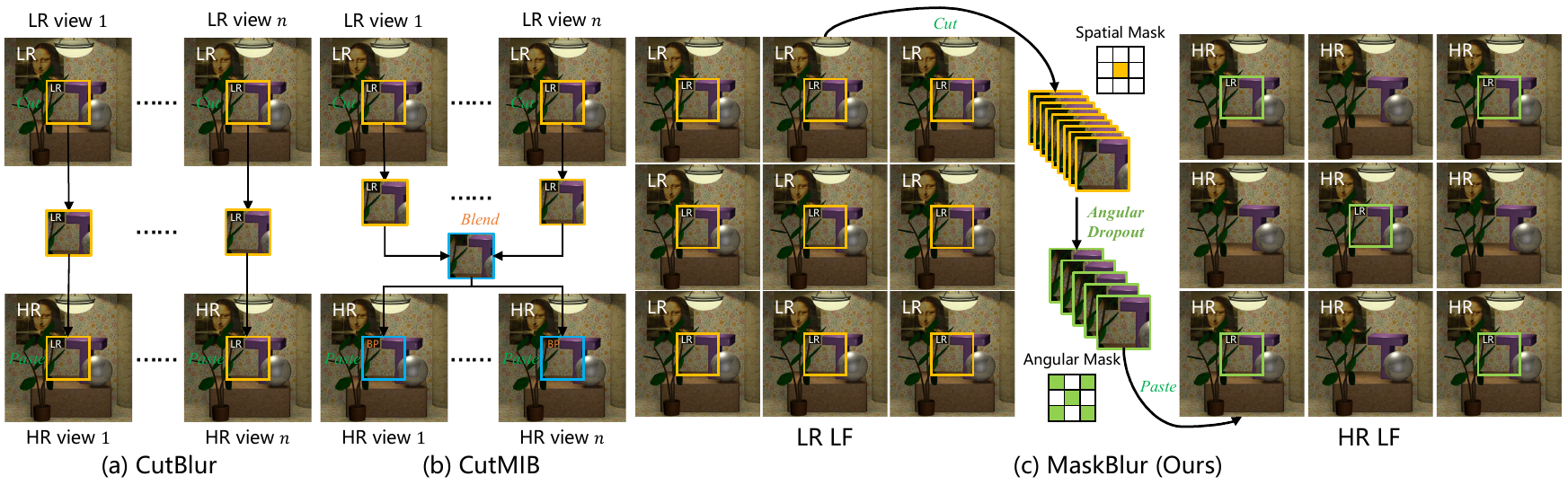}
  \caption{Detailed comparisons of MaskBlur with CutBlur~\cite{yoo2020rethinking} and CutMIB~\cite{xiao2023cutmib}. MaskBlur consists of spatial blur and angular dropout, which are implemented through the spatial and angular masks, respectively.}
  \label{fig:intro_mb}
\end{figure*}

\begin{figure}[tb]
  \centering
  \includegraphics[width=\linewidth]{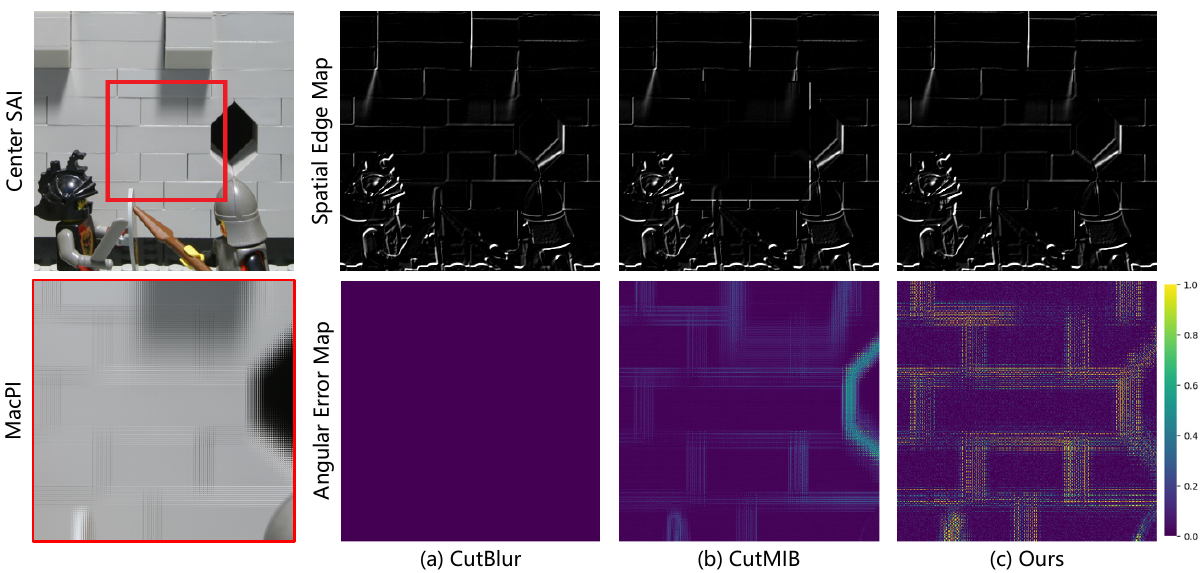}
  \caption{Analyzing CutBlur~\cite{yoo2020rethinking}, CutMIB~\cite{xiao2023cutmib} and MaskBlur from spatial and angular domain. The center SAI is in a 5 $\times$ 5 LF. The red rectangle denotes the area for the cutting and pasting operation, i.e., cut an LR patch and paste it to the original HR image.
  Macro-pixel image (MacPI) represents the pasted region in the angular domain. 
  The spatial edge map is generated by using \textit{Sobel} operation on the augmented HR LF image the angular error map is calculated between LR and augmented HR LF in the angular domain.
  MaskBlur can maintain spatial structure consistency and enhance angular domain information. Please zoom in for better visualization.}
  \label{fig:intro_comp}
\end{figure}

Many previous methods on DA primarily focus on high-level vision tasks such as image classification~\cite{he2019bag} and object detection~\cite{zhang2019bag}. However, few studies have been performed on DA specifically for LF image SR. The commonly used DA techniques for LF image SR tasks still mainly rely on geometric transformations, e.g., horizontal and vertical flips. 
A recent study in ~\cite{yoo2020rethinking} demonstrated the importance of maintaining spatial structure consistency for image SR, i.e., low-resolution (LR) and high-resolution (HR) have the same structure and further introduced the CutBlur DA strategy by randomly exchanging pixels between LR and the corresponding HR regions through a \textit{cutting-pasting} operation. 
While a direct extension of CutBlur from general images to LF images is a straightforward approach, i.e., applying CutBlur to each LF sub-aperture image (SAI), as shown in Fig.~\ref{fig:intro_mb} (a). However, this approach only maintains spatial structure consistency of the LF image while neglecting to augment the angular domain, and the latter has been validated to be crucial for LF image SR works \cite{jin2020light,wang2022disentangling}. 
A subsequent study~\cite{xiao2023cutmib} proposed CutMIB that implicitly utilizes the angular (multi-view) information of the LF image by involving a \textit{cutting-blending-pasting} operation, e.g., cropping patches of each LR SAI, blending them, i.e., calculate the mean of these patches, and pasting the blended patch onto corresponding HR SAIs, as shown in Fig.~\ref{fig:intro_mb} (b). 
However, CutMIB implicitly enhances the angular information of the LF image, it does not consider the differences between individual SAIs and disrupts spatial structure consistency, thereby limiting its performance.

To better demonstrate our observations, we provide visual comparisons using the spatial edge map, which contains structure information, and the angular error map to represent angular domain information difference in Fig.~\ref{fig:intro_comp}. Specifically, we use CutBlur or CutMIB on the LR center view in a 5$\times$5 LF, cut an LR patch or cut and blend an LR patch, and then paste it to the original HR image. Finally, we adopt \textit{Sobel} operation on augmented HR LF image to generate the spatial edge map and calculate the angular error map between LR and augmented HR LF in the angular domain. We can directly observe from the spatial edge map in Fig.~\ref{fig:intro_comp}(a) that the CutBlur strategy has the complete edge in the pasted patch region which can represent it maintains spatial structure consistency, but it does not augment the information of angular domain, i.e., the information of pasted region all comes from LR domain (see angular error map in Fig.~\ref{fig:intro_comp}(a)). 
Then, we can see from the angular error map in Fig.~\ref{fig:intro_comp}(b) that the CutMIB strategy provides additional angular (multi-view) information in the pasted region. However, the spatial edge map in Fig.~\ref{fig:intro_comp}(b) shows that the CutMIB has obvious edge differences adjacent to the pasted region, and the original edge information is missing inside the region i.e., it cannot maintain spatial structure consistency.

To simultaneously augment both spatial domain and angular domain information based on the observations above, this paper proposes a novel DA approach called MaskBlur for LF image SR (see Fig.~\ref{fig:intro_mb} (c)). The spatial edge map and angular error map in Fig.~\ref{fig:intro_comp} (c) shows that our method not only preserves structure consistency in the spatial domain (i.e., maintaining the same edge structure) but also incorporates angular domain information from distinct LR and HR spaces.
Specifically, MaskBlur comprises spatial blur and angular dropout, which are responsible for enhancing spatial and angular information, respectively. Firstly, spatial blur determines which regions of the LR and HR images need pixel swapping by randomly generating a spatial mask. Secondly, angular dropout selects which views are subjected to spatial blur operations based on a randomly generated angular mask. MaskBlur enables the LF image SR model to treat pixels in the spatial and angular domains differently, thus improving model regularization. Fig.~\ref{fig:intro_chart} illustrates that MaskBlur improves the values of PSNR by a large margin compared to other DA schemes on various LF image SR methods while maintaining the network structures unchanged. Moreover, we verify the effectiveness of the  MaskBlur on real-world LF image SR, denoising, and deblurring tasks.

In summary, the contributions of this paper are as follows:
\begin{itemize}
\item This paper proposes MaskBlur, a simple yet effective DA scheme for LF image SR. MaskBlur is designed to preserve spatial structure consistency while simultaneously enhancing angular domain information.

\item MaskBlur comprises spatial blur and angular dropout two key components. Spatial blur determines which regions of the LR and HR images need pixel swapping by randomly generating a spatial mask. Angular dropout selects which views undergo spatial blur operations based on a randomly generated angular mask.

\item The efficacy of MaskBlur is substantiated through comprehensive experiments across diverse LF image SR methods. Moreover, MaskBlur can be extended to other LF image processing tasks, such as LF image denoising, deblurring, and real-world SR.
% For example, the 4$\times$ SR task significantly improves the results of existing methods by 0.1~0.3 dB. 
% MaskBlur can also be applied to 
\end{itemize}

\section{Related Work}
\label{sec: related}

This section briefly reviews the related works of the LF image SR task and the DA in computer vision.

\subsection{LF image SR.}
LF image SR aims to generate HR LF images from LR LF inputs. One straightforward approach for LF image SR is to independently apply single-image super-resolution (SISR) methods to each SAI. However, directly applying SISR methods to LF images may not produce satisfactory results as it overlooks the correlations between different viewpoints. To achieve high-performance LF image SR, it is important to fully exploit the information within a single viewpoint (i.e., spatial information) and across different viewpoints (i.e., angular information).

\noindent \textbf{Traditional Methods.} Traditional methods for LF image SR relied on the geometric structure \cite{liang2015light} and mathematical modeling \cite{wanner2013variational} of the LF, which utilized projection and optimization techniques to super-resolve LR images. 
Bishop \etal \cite{bishop2011light} first restores the scene depth, and then uses the Bayesian deconvolution method to obtain HR results.
Mitra \etal \cite{mitra2012light} introduced a Gaussian mixture model to encode the structure of the LF for LF image SR. 
Wanner \etal \cite{wanner2013variational} utilized the structure tensor to deduce depth from Epipolar Plane Images (EPIs), employing this data to refine the resolution of view images through interpolation.
Farrugia \etal \cite{farrugia2017super} decomposed HR and LR image patches into subspaces and proposed a linear subspace projection method. 
Alain \etal \cite{alain2018light} extended the BM3D \cite{egiazarian2015single} filter to LFBM5D for LF denoising and spatial SR. 
Rossi \etal \cite{rossi2018geometry} developed a graph-based approach to achieve spatial super-resolution through graph optimization.
However, these methods are computationally intensive and have lower accuracy.

\noindent \textbf{Deep learning-based Methods.}
In recent years, CNN-based \cite{yoon2017light} and Transformer-based methods \cite{liang2022light,wang2022detail,liang2023learning} have become mainstream, demonstrating significant improvements in speed and accuracy compared to traditional approaches.
LFCNN \cite{yoon2017light} was the first CNN-based method that learns the correspondence between the LF stacked sub-aperture images (SAIs). 
Jin \etal \cite{jin2020light} proposed LFSSR-ATO, a \textit{all-to-one} LF image SR method, which incorporates structural consistency regularization to preserve disparity structures.
Wang \etal \cite{wang2020spatial} proposed an LF-InterNet interact spatial and angular information for LF image SR.
Zhang \etal \cite{zhang2019residual} divided the LF into four different branches based on specific angular directions and utilized four sub-networks to model the spatial-angular correlations. 
Liu \etal \cite{liu2021intra} proposed LF-IINet to exploit the correlations among all views and simultaneously preserve the parallax structure of LF views.
Wang \etal \cite{wang2022disentangling} disentangled the spatial-angular correlation and proposed the method DistgSSR for LF image SR. 
Liang \etal \cite{liang2022light} applied the Transformer to LF image SR for the first time and designed a Spatial Transformer and Angular Transformer to incorporate spatial and angular information respectively. 
Wang \etal \cite{wang2022detail} treated LF image SR as a sequence-to-sequence reconstruction task and designed a DPT to leverage gradient maps of LF to guide the sequence learning.
While existing methods concentrate on optimizing the utilization of spatial and angular information in LF images, their performances frequently face limitations stemming from the size of the LF dataset. More recently, EPIT \cite{liang2023learning} has delved further into addressing the challenges posed by large disparity variations inherent in LF. HLFSR~\cite{van2023light} proposed an inter-intra spatial feature extractor and an inter-intra angular feature extractor to leverage the correlations among pixels in the spatial-angular domain. LF-DET~\cite{cong2023exploiting} introduced a deep efficient transformer architecture with a spatial-angular separable transformer encoder. It incorporates sub-sample spatial modeling and multi-scale angular modeling to facilitate global context interaction.

We focus on an orthogonal direction, i.e., the perspective of DA. The proposed approach, MaskBlur, offers generalization and can be applied to different LF image SR methods effectively.

\subsection{DA in Computer Vision}

\noindent \textbf{DA in high-level vision.} DA techniques have been widely studied for high-level vision tasks, such as image classification \cite{zhang2019bag}. These techniques are primarily implemented in pixel space and feature space. In pixel space, general DA strategies include geometric and color transformations, i.e., translation, rotation, flipping (horizontal or vertical), color jitter, RGB permute, and blending. Mixup \cite{zhang2017mixup} randomly selected two images and blended them in proportion to generate unseen training samples. Cutout \cite{devries2017improved} randomly removed selected regions in an image. CutMix \cite{yun2019cutmix} combined the advantages of Mixup and Cutout by replacing randomly removed regions with content from another image. 
Hide-and-seek \cite{kumar2017hide} divided an image into grids and randomly hid image blocks for model training.
AutoAugment\cite{cubuk2018autoaugment} focused on automatically learning the optimal augmentation strategy. 
Recently, BEiT\cite{bao2021beit} and MAE\cite{he2022masked} have used the random masking strategy to restore the masked tokens or pixels in a self-supervised manner for pre-training image transformers.
In feature space, some methods are specifically designed to enhance the features of CNNs, including feature mixing, shaking, and dropping. Feature mixing is similar to Mixup, while Manifold Mixup \cite{verma2019manifold} blended the features of two images in feature space. Shake-shake introduced multiple branches in the network and performed random shaking operations in each branch. Shake-Shake \cite{gastaldi2017shake} and ShakeDrop \cite{yamada2019shakedrop} introduced random pruning operations. DropPath \cite{larsson2016fractalnet} randomly drooped certain connections in the network during training. However, these methods are primarily designed for high-level vision tasks and may lose spatial structure information, which is important for low-level vision tasks, such as SR and denoising.

\noindent \textbf{DA in low-level vision.} 
\cite{timofte2016seven} proposed seven techniques to improve the performance of single-image SR, including pixel geometric transformations such as horizontal flipping, vertical flipping, and rotation. 
\cite{feng2019suppressing} analyzes the application of Mixup in single-image SR to mitigate model overfitting.
CutBlur~\cite{yoo2020rethinking} comprehensively analyzed existing DA methods for SR tasks, highlighted the importance of maintaining spatial structure consistency, and proposed the \textit{cutting-pasting} operation to exchange information between LR and HR images to generate new training samples. However, the CutBlur method mainly operates in the spatial domain and does not consider the angular information when LF image SR. Recently, \cite{xiao2023cutmib} introduced a DA strategy CutMIB for LF image SR task which leverages multi-view information during the training phase. CutMIB involves a \textit{cutting-blending-pasting} operation, i.e., cropping and blending random regions of each LR SAI and pasting them onto corresponding HR SAIs. However, while CutMIB implicitly utilizes the multi-view information of LF images, it does not consider the differences between individual SAIs and may disrupt spatial structure consistency. 

In this study, we propose MaskBlur, a DA strategy designed to enhance spatial and angular information through spatial blur and angular dropout. Our approach not only preserves spatial structure consistency but also effectively enhances the angular information of LF images.

\begin{figure*}[tb]
  \centering
  \includegraphics[width=\linewidth]{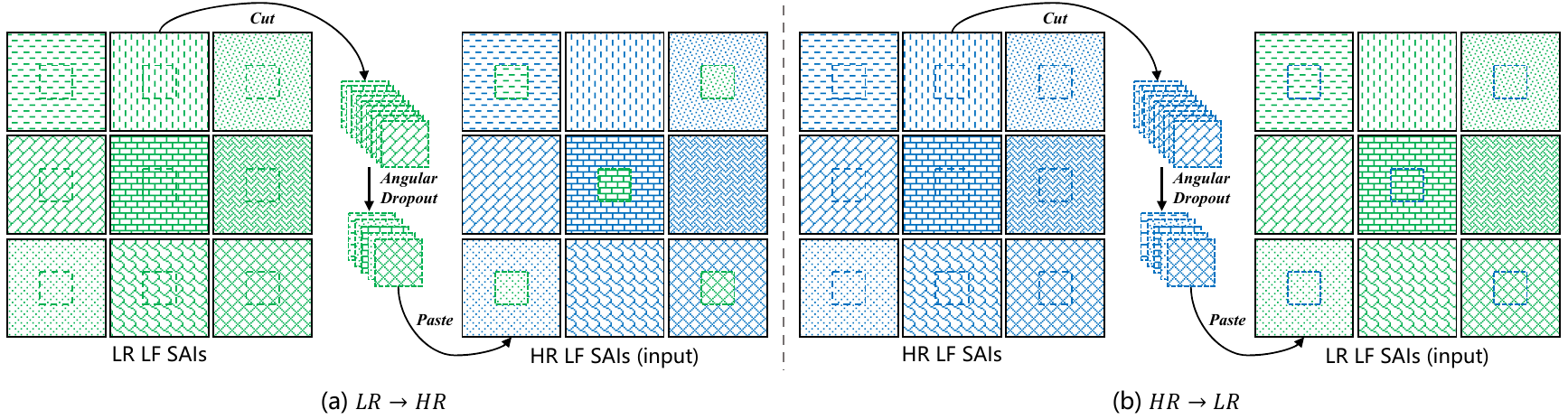}
  \caption{Semantic illustration of MaskBlur. Here, an LF of size $U=V=3, H=W=3$ is used as a toy example. LR LF SAIs and HR LF SAIs are painted with \protect \textcolor{green}{green} and \textcolor{blue}{blue} background color. Different textures denote different views. We show the examples of $LR \rightarrow HR$ and $HR \rightarrow LR$. Better to view in color and texture.}
  \label{fig:method_mb}
\end{figure*}

\section{Method}
\label{sec: method}

\subsection{Preliminary}
We adopt the two-plane LF parameterization model \cite{levoy1996light} to represent an LF image, which can be formulated as a 4D function $\mathcal{L} (u,v,h,w)\in \mathbb{R}^{U\times V\times H\times W}$, where $U$ and $V$ stand for angular dimensions, $H$ and $W$ denote spatial dimensions. 
LF image SR networks aim to learn the mapping function $f_\theta(\cdot)$ and take an LR LF $\mathcal{L}_{LR}\in \mathbb{R}^{U\times V\times H\times W}$ as input and generate the SR LF $\mathcal{L}_{SR}\in \mathbb{R}^{U\times V\times \alpha H\times \alpha W}$ as output, where $\alpha$ stands for the scale factor of SR, which can be denoted as $\mathcal{L}_{SR} = f_\theta(\mathcal{L}_{LR})$, where $\theta$ is the parameters of the LF image SR network. The reconstructed SR LF is desired to be close to the ground-truth HR LF $\mathcal{L}_{HR}\in \mathbb{R}^{U\times V\times \alpha H\times \alpha W}$. The LF image SR network is optimized with a loss function, and $L_1$ is the most commonly used one. Given a training set {$\mathcal{L}^{LR}_i, \mathcal{L}^{HR}_i$}, which contains $N$ LR input LF and HR counterpart. The goal of training the LF image SR network is to minimize the $L_1$ loss function:

\begin{equation}
\begin{aligned}
L=\frac{1}{N}\sum_{i=1}^{N}\left \| f_\theta(\mathcal{L}^{LR}_i)-\mathcal{L}^{HR}_i \right \| _1  
\end{aligned}
\end{equation}

% In our experiments, we set the $\alpha = 4$ to evaluate the impact of DA on various models' performance. 
In our experiments, we set $\alpha=4$ to evaluate the impact of DA on various models’ performance. Most mainstream LFSR methods opt for a scaling factor of $\alpha=4$ for 4$\times$ super-resolution tasks. This choice provides a reliable baseline model and facilitates the comparison of different DA schemes against the baseline. Additionally, the 4$\times$ super-resolution task is more challenging than the 2$\times$, better reflecting the impact of DA schemes. Therefore, we ultimately chose $\alpha=4$ for 4$\times$ super-resolution tasks.

\subsection{Motivation}
Currently, common DA techniques used in LF image SR tasks mainly involve geometric transformations, such as horizontal or vertical flips. However, these geometric transformations do not consider the characteristics of the SR task, where images may contain varying degrees of blurriness that need to be treated differently by the model. Therefore, the effectiveness is limited.
CutBlur~\cite{yoo2020rethinking} was specifically designed for the single SR task and could maintain the spatial structure consistency of the augmented image through the \textit{cutting-pasting} operation. 
However, extending CutBlur directly from single images to LF images does not enhance the angular domain information, i.e., all the information comes from the LR or HR image when extracting the angular domain features.
Subsequent work, CutMIB~\cite{xiao2023cutmib}, utilized multi-view information and implicitly enhances the angular domain (multi-view) information through \textit{cutting-blending-pasting} operations. However, all the pasted regions from different viewpoints contain the same spatial information, which violates the spatial structure consistency.
LF images consist of both spatial and angular information. Therefore, we think that a better DA technique is needed to augment simultaneously LF images in both spatial and angular domains. Specifically, we extend the CutBlur operation to spatial blur in the spatial domain. We control the position of the pasted region using a patch-based random spatial mask instead of a rectangle mask. In the angular domain, drawing inspiration from the commonly used regularization technique Dropout~\cite{srivastava2014dropout}, we introduce an additional operation called Angular Dropout. By using a random angular mask, we perform random selections on the viewpoints to control whether the spatial blur operation is applied. The pasted region in the angular domain will contain information from both the LR and HR images, thereby enhancing the angular domain information. The specific details of MaskBlur are described in Sec.~\ref{maskblur}.

\subsection{MaskBlur}
\label{maskblur}

As shown in Fig. \ref{fig:method_mb}, our MaskBlur consists of spatial blur and angular dropout to generate new training samples.

\noindent \textbf{Spatial Blur} randomly generate a spatial mask $\mathrm{M_{spa}}$ for $k$-th SAI from $\mathcal{L}_{LR}$ and utilize the spatial blur operation between $\mathcal{L}_{LR}$ and $\mathcal{L}_{HR}$.

\begin{equation}
\begin{aligned}
& \hat{\mathcal{L}}_{k}^{LR\to HR} = \mathrm{M_{spa}}\odot \mathcal{L}_{k}^{LR} +(\mathbf{1} -\mathrm{M_{spa}})\odot \mathcal{L}_{k}^{HR \downarrow s}\\
& \hat{\mathcal{L}}_{k}^{HR\to LR} = \mathrm{M_{spa}}\odot \mathcal{L}_{k}^{HR \downarrow s}+(\mathbf{1} -\mathrm{M_{spa}})\odot\mathcal{L}_{k}^{LR}
\end{aligned}
\end{equation}
where $\mathcal{L}_{k}^{HR \downarrow s}$ is achieved by downsampling the $s$ ratio based on bilinear interpolation, $\mathrm{M_{spa}} \in\left \{ 0,1 \right \} ^{H\times W}$ denotes a binary mask indicating where to mask, $\mathbf{1}$ is a binary mask filled with ones, and $\odot$ is the element-wise multiplication. 
% Specifically, when generating the spatial mask, we find that using image blocks (or patches) instead of pixels as the minimum unit of the mask can get better results, as it can preserve the local completeness of the information. Detailed experiments are shown in Sec.~\ref{sec: abla}.

\noindent \textbf{Angular Dropout} randomly generate an angular mask $\mathrm{M_{ang}}$ for all views of LF and carry out the angular dropout operation.

\begin{equation}
\begin{aligned}
& \hat{\mathcal{L}}_{k}^{LR\to HR} = 
\begin{cases}
  \hat{\mathcal{L}}_{k}^{LR\to HR}, \text{ if } \mathrm{M_{ang}}[k]==1 \\
  \mathcal{L}_{k}^{LR}, \text{ if } \mathrm{M_{ang}}[k]==0
\end{cases}
, k\in[1,UV] \\
& \hat{\mathcal{L}}_{k}^{HR\to LR} = 
\begin{cases}
  \hat{\mathcal{L}}_{k}^{HR\to LR}, \text{ if } \mathrm{M_{ang}}[k]==1 \\
  \mathcal{L}_{k}^{LR}, \text{ if } \mathrm{M_{ang}}[k]==0
\end{cases}
, k\in[1,UV]
\end{aligned}
\end{equation}
where $\mathrm{M_{ang}} \in\left \{ 0,1 \right \} ^{UV}$ denotes an angular binary mask representing which views are selected to perform the spatial blur, and $UV$ indicates the angular number. Therefore, we can generate a new pair training sample $\left \{\hat{\mathcal{L}}^{LR\to HR},\hat{\mathcal{L}}^{HR\to LR} \right \}$ through the spatial blur and angular dropout operations for each SAI from $\mathcal{L}_{LR}$. The procedure of MaskBlur is summarized in Algorithm~\ref{alg:algorithm}. Note that our MaskBlur is a more generalized form than the CutBlur. When the spatial mask forms a single continuous rectangle and the angular mask is filled with ones, the MaskBlur is equivalent to the CutBlur.

\begin{algorithm}[tb]
    \caption{MaskBlur algorithm}
    \label{alg:algorithm}
    \textbf{Input}: LR LF image patch $\mathcal{L}_{LR}$, HR LF image patch $\mathcal{L}_{HR}$, DA probability $p$, spatial mask ratio $s_r$, spatial mask block-wise size $s_{size}$, angular mask ratio $a_r$\\
    \textbf{Output}: $\mathcal{L}_{LR}$, $\mathcal{L}_{HR}$
    \begin{algorithmic}[1] %[1] enables line numbers
        \STATE  $\mathrm{M_{spa}} = \mathrm{spa\_mask}(s_r, s_{size})$.
        \STATE  $\mathrm{M_{ang}} = \mathrm{ang\_mask}(a_r).\mathrm{reshape}(1,UV)$.
        % \STATE  $\mathrm{Index_{ang}} = \mathrm{M_{ang}}==1$.
        \IF {$\mathrm{Rand}(0,1)<p$}
            \FOR{$k=1$ to $UV$}
                \STATE  // Angular Dropout
                \IF {$\mathrm{M_{ang}}[k]==1$} 
                    \STATE  // Spatial Blur
                    \IF {$\mathrm{Rand}(0,1)>0.5$}
                    \STATE  $\mathcal{L}_{k}^{LR} = \mathrm{M_{spa}}\odot \mathcal{L}_{k}^{LR} +(\mathbf{1} -\mathrm{M_{spa}})\odot \mathcal{L}_{k}^{HR \downarrow s}$
                    \STATE \COMMENT{$LR\to HR$}
                    \ELSE
                    \STATE  $\mathcal{L}_{k}^{LR} = \mathrm{M_{spa}}\odot \mathcal{L}_{k}^{HR \downarrow s}+(\mathbf{1} -\mathrm{M_{spa}})\odot\mathcal{L}_{k}^{LR}$
                    \STATE \COMMENT{$HR\to LR$}
                    \ENDIF
                \ENDIF
            \ENDFOR
        \ENDIF
        \STATE \textbf{return} $\mathcal{L}_{LR}$, $\mathcal{L}_{HR}$
    \end{algorithmic}
\end{algorithm}

\begin{figure}[tb]
  \centering
  \includegraphics[width= \linewidth]{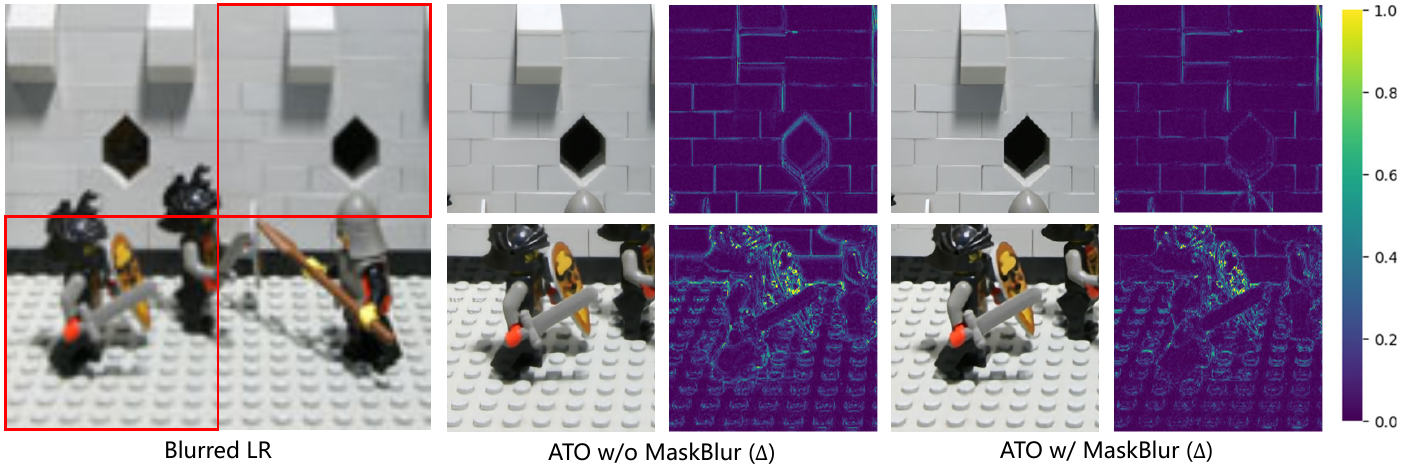}
  \caption{Qualitative comparison of the ATO~\cite{jin2020light} with and without MaskBlur when the network inputs the mask blurred image during the inference. Red boxes indicate the regions of the spatial mask. $\Delta$ is the absolute residual intensity map between the network output and the ground-truth HR image. MaskBlur successfully generates clear results in more blurred areas (red boxes) while the ATO generates unrealistic artifacts.}
  \label{fig:inpt_da_compare}
\end{figure}

\subsection{Further Visual Analysis}

There may be various degrees of blur in the real LF image. MaskBlur aims to enable the model to treat pixels differently when super-resolving LF images rather than blindly treating all pixels equally. To verify this, we perform the mask blur on some regions of the input LR image. We input blurred images into ATO~\cite{jin2020light} and ATO-Maskblur models respectively, and the results are shown in Fig.~\ref{fig:inpt_da_compare}. The residual intensity map has significantly decreased in the model trained with MaskBlur. 
When the SR model takes the blurred image at the test phase, it commonly produces overly blurred predictions, especially in areas with lines. MaskBlur can resolve this issue by directly providing such examples to the model during the training phase. Therefore, the models trained with MaskBlur enable discriminatively applying SR to the image. 

\subsection{Discussion}

MaskBlur consists of spatial blur and angular dropout controlled by spatial and angular masks, respectively. Naturally, we need to discuss how to generate spatial masks and angular masks. For the spatial mask, we need to consider which mask sampling strategy to use, e.g., rectangular sampling~\cite{yoo2020rethinking,xiao2023cutmib}, grid sampling, or random sampling~\cite{kumar2017hide,he2022masked}, as shown in Fig.~\ref{fig:spa_mask}. Additionally, we also need to discuss the impact of block-wise size on the spatial mask. Specifically, when generating the spatial mask, we find that using image blocks (or patches) instead of pixels as the minimum unit of the mask can yield better results as it preserves the local completeness of the information. Regarding the angular mask, we continue to consider which sampling strategy to use, e.g., grid or random. LF SAIs are stacked into different view stacks along four specific directions, namely horizontal, vertical, main diagonal, and anti-diagonal. These specific directions capture the implicit geometric information of the LF image, which has been proven to be crucial in tasks like LF depth estimation~\cite{shin2018epinet,huang2021fast} and SR~\cite{zhang2019residual}.
Therefore, we also explore the performance of the angular mask on SAIs in different directions and their combinations (see Fig.~\ref{fig:ang_mask}). Angular dropout can also be combined with other spatial data augmentation strategies to further improve performance. Furthermore, we investigate the probabilities of MaskBlur, spatial mask, and angular mask, as well as the performance of MaskBlur with different dataset sizes. Detailed experimental results and analyses are presented in Sec.~\ref{sec: abla}.

\section{Experiments}
\label{sec: experi}

\subsection{Experimental Settings}
\label{sec: setting}

\noindent \textbf{Dataset and evaluation.} Following the experimental settings of previous works \cite{wang2020spatial,cong2023exploiting}, we utilize five prevalent LF image datasets: EPFL \cite{rerabek2016new}, HCInew \cite{honauer2016dataset}, HCIold \cite{wanner2013datasets}, INRIA \cite{le2018light}, and STFgantry \cite{vaish2008new}, for both model training and testing based on BasicLFSR\footnote{\url{https://github.com/ZhengyuLiang24/BasicLFSR}}. Specifically, the angular resolution of each LF image is 9$\times$9, and we extract the central 5$\times$5 SAIs for performing the SR task. The LF images are converted to the YCbCr color space, and the model performs SR on the Y channel. The Cb and Cr channels are upsampled using bicubic interpolation. For evaluation, we employ the PSNR and SSIM~\cite{wang2004image} metrics on the Y channel as the main metric on several benchmarks. Each dataset consists of multiple test scenes. We first calculate the metric scores for each scene by averaging the scores of all SAIs individually. Then, we compute the average score for the dataset by averaging the scores across multiple scenes. Finally, we obtain the overall average score by averaging the scores across all datasets.

\noindent \textbf{Implementation details.} We select five well-known LF image SR models, i.e., InterNet \cite{wang2020spatial}, ATO \cite{jin2020light}, IINet \cite{liu2021intra}, DistgSSR~\cite{wang2022disentangling} and EPIT~\cite{liang2023learning} as baseline models to evaluate the generalization of MaskBlur, since these models differ significantly in terms of architecture and size. 
% DPT\cite{wang2022detail}
For a fair comparison, we train the baseline models with and without MaskBlur from scratch, following the recommended configurations provided by the authors' code. During training, we randomly crop LF images into 128$\times$128 image patches and downsample these patches using bicubic interpolation with a scale factor of 4, resulting in LR LF patches of size 32$\times$32, which serve as inputs to the 4$\times$ SR models. We employ the Adam optimizer with a batch size of 8 and a learning rate of 2e-4, which is halved every 15 epochs. The default DA strategy is conducted by random 90-degree rotation and flipping in horizontal and vertical directions. The training of these baseline models is implemented in PyTorch and conducted on an NVIDIA A100 GPU. 
% Please refer to the \textbf{supplementary material} for more details and experimental results.

\begin{table}[tb]
    \centering
    \caption{Spatial blur ablation experiments with InterNet \protect\cite{wang2020spatial} in 4$\times$ SR task on benchmark datasets. We report the average PSNR (dB, $\uparrow$) and $\delta$ indicates the performance gap from the baseline. Default settings are marked in \colorbox {gray!30}{gray}.}
    \begin{tabular}{lllc}
        \toprule
        Spa. mask samp.  & ratio & block & PSNR ($\delta$) \\
        \midrule
        InterNet-base      & - & - & 31.596 (+0.000)      \\
        \midrule
        rectangle    & 50\% & 4 & 31.635 \green{(+0.039)}       \\
        grid         & 50\% & 4 & 31.640 \green{(+0.044)}       \\
        learning     & auto  & 4 & \textbf{31.653} \green{(+0.057)} \\
        \rowcolor{gray!30} random       & 50\% & 4 & 31.651 \green{(+0.055)}  \\
        \midrule
        random         & 50\% & 1 & 31.623 \green{(+0.027)}       \\
        random         & 50\% & 2 & 31.626 \green{(+0.030)}       \\
        \rowcolor{gray!30} random         & 50\% & 4 & \textbf{31.651 \green{(+0.055)}} \\
        random         & 50\% & 8 & 31.636 \green{(+0.040)}       \\
        random         & 50\% & 16 & 31.628 \green{(+0.032)}       \\
        \midrule
        random         & 25\% & 4 & 31.633 \green{(+0.037)}       \\
        \rowcolor{gray!30} random         & 50\% & 4 &  \textbf{31.651 \green{(+0.055)}} \\
        % \cellcolor{gray!30}
        random         & 75\% & 4 & 31.602 \green{(+0.006)}       \\
        \bottomrule
    \end{tabular}
    
    \label{tab:abl_spa}
\end{table}

\begin{figure}[tb]
  \centering
  \includegraphics[width=0.75 \linewidth]{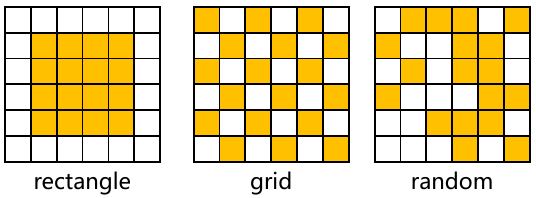}
  \caption{Illustration of three different spatial mask sampling strategies: rectangle, grid, and random (default).}
  \label{fig:spa_mask}
\end{figure}

\subsection{Ablation Study}
\label{sec: abla}

% \subsubsection{Spatial mask strategies.}
% \label{sec: spa_strategy}

\noindent \textbf{Spatial mask sampling strategy.} Given a masking ratio of 0.5, we compare three different mask sampling strategies:  rectangle sampling, grid sampling,  and random sampling, as illustrated in Fig. \ref{fig:spa_mask}. Table \ref{tab:abl_spa} presents the comparative results of these sampling strategies, indicating that all three sampling strategies significantly outperform the baseline, demonstrating the effectiveness of spatial blur. Among them, random sampling performs the best, and we conduct further analysis. 
Rectangle sampling leads to a significant loss of information in the sampled central region, making it challenging for the model to learn effectively. Grid sampling performs better than rectangle sampling due to reduced information loss around each patch. However, due to the fixed sampling patches, its representational quality is relatively low. Random sampling involves randomly selecting image patches to form the mask region, which not only has less information loss but also has the strongest representation. It serves as a general sampling strategy, with rectangle sampling and grid sampling as special cases. We also conducted a comparative analysis of the random sampling strategy and the learning-based approach. Specifically, we employ two convolutional layers with sigmoid activation to automatically generate the spatial mask for spatial blur. The results indicate that while the learning-based approach offers minimal improvement over the random sampling strategy, it introduces additional parameters and computational effort, slowing the training speed by about three minutes per epoch compared to the random sampling strategy. In contrast, the InterNet model trained with the random sampling strategy only experiences an increase of approximately one minute per epoch. Therefore, the random sampling strategy provides a better trade-off between performance and computational efficiency.

\begin{table}[tb]
    \centering
    % \caption{
    %  Angular dropout ablation experiments with InterNet in 4$\times$ SR task on benchmark datasets.}
    \caption{
    Angular dropout ablation experiments with InterNet \protect\cite{wang2020spatial} in 4$\times$ SR task on benchmark datasets. We report the average PSNR (dB, $\uparrow$) and $\delta$ indicates the performance gap from the baseline. Default settings are marked in \colorbox {gray!30}{gray}.
    }
    \begin{tabular}{llc}
        \toprule
        Angular mask sampling  & ratio & PSNR ($\delta$) \\
        \midrule
        InterNet-base       & - & 31.596 (+0.000)      \\
        InterNet+spatial blur       & - & 31.651 \green{(+0.055)}      \\
        \midrule
        grid         & 76\% & 31.658 \green{(+0.062)}       \\
        % \red{learning}     & \red{auto} & \red{31.658 \green{(+0.062)}} \\
        % \rowcolor{gray!30} random       & 75\% & \cellcolor{gray!30} \textbf{31.732 \green{(+0.136)}}  \\
        % \midrule
        horizontal       & 80\% & 31.656 \green{(+0.060)}       \\
        vertical         & 80\% & 31.661 \green{(+0.065)}       \\
        main diagonal         & 80\% & 31.658 \green{(+0.062)}       \\
        antidiagonal         & 80\% & 31.668 \green{(+0.072)}       \\
        cross           & 64\% & 31.668 \green{(+0.072)}             \\
        bidiagonal & 64\% &    31.669 \green{(+0.073)}     \\
        cross+bidiagonal & 32\% & 31.671 \green{(+0.075)}       \\
        \rowcolor{gray!30} random       & 75\% & \textbf{31.732 \green{(+0.136)}}  \\
        \midrule
        random       & 25\% & 31.676 \green{(+0.080)}       \\
        random       & 50\% & 31.703 \green{(+0.097)}       \\
        \rowcolor{gray!30} random       & 75\% &  \textbf{31.732 \green{(+0.136)}}  \\
        \bottomrule
    \end{tabular}
    
    \label{tab:abl_ang}
\end{table}

\begin{figure}[t]
  \centering
  \includegraphics[width=\linewidth]{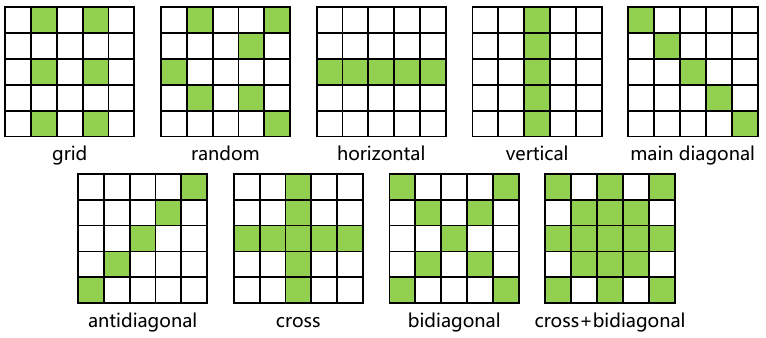}
  \caption{Illustration of different angular mask sampling strategies: rectangle, random (default), and special direction and combinations.}
  \label{fig:ang_mask}
\end{figure}

\begin{table}[t]
    \centering
    \caption{Angular dropout combined with other spatial DA strategies with InterNet \protect\cite{wang2020spatial} in 4$\times$ SR task on benchmark datasets. $\delta$ indicates the performance gap from the baseline.}
    \begin{tabular}{lc}
        \toprule
        DA strategy         & $\delta$ \\
        \midrule
        RGB permute  &  \green{+0.009}       \\
        Cutout       &  \green{+0.001}       \\
        Mixup        &  \green{+0.025}       \\
        CutBlur      &  \green{+0.033}       \\
        saptial blur      & \green{+0.055}       \\
        \midrule
        RGB permute+angular dropout   & \green{+0.042}       \\
        Cutout+angular dropout        & \green{+0.049}       \\
        Mixup+angular dropout         & \green{+0.045}       \\
        CutBlur+angular dropout       & \green{+0.087}       \\
        \rowcolor{gray!30} spatial blur+angular dropout (MaskBlur)   & \textbf{\green{+0.136}}       \\
        \bottomrule
    \end{tabular}
    \label{tab:abl_comb}
\end{table}

\begin{table}[t]
    \centering
    
    \caption{
     DA probability $p$ ablation experiments with InterNet~\cite{wang2020spatial} in 4$\times$ SR task on benchmark datasets. We report the average PSNR (dB, $\uparrow$) and $\delta$ indicates the performance gap from the baseline. Default settings are marked in \colorbox {gray!30}{gray}.}
    % \renewcommand\arraystretch{1.0}
    % \resizebox{0.6 \columnwidth}{!}{
    \begin{tabular}{lcc}
        \toprule
        Method  & $p$ & PSNR ($\delta$) \\
        \midrule
        InterNet-base      & 0.00 & 31.596 (+0.000)      \\
        \rowcolor{gray!30} +MaskBlur  & 0.25 &  \textbf{31.732 \green{(+0.136)}}      \\
        +MaskBlur  & 0.50 & 31.697 \green{(+0.101)}      \\
        +MaskBlur  & 0.75 & 31.652 \green{(+0.056)}      \\
        +MaskBlur  & 1.00 & 31.679 \green{(+0.083)}      \\
        \bottomrule
    \end{tabular}
    % }
    
    \label{tab:abl_prob}
\end{table}

\begin{table}[t]
    \centering
    \caption{
     Various dataset sizes ablation experiments with InterNet \cite{wang2020spatial} in 4$\times$ SR task on benchmark datasets. We report the average PSNR (dB, $\uparrow$) and $\delta$ indicates the performance gap from the baseline. Default settings are marked in \colorbox {gray!30}{gray}.}
    \begin{tabular}{llc}
        \toprule
        Method  & Data Size & PSNR ($\delta$) \\
        \midrule
        InterNet-base      & 25\% & 30.903 (+0.000)      \\
        \rowcolor{gray!30} +MaskBlur  & 25\% &  \textbf{31.104 \green{(+0.201)}}      \\
        \midrule
        InterNet-base      & 50\% & 31.175 (+0.000)      \\
        \rowcolor{gray!30} +MaskBlur  & 50\% &  \textbf{31.349 \green{(+0.174)}}      \\
        \midrule
        InterNet-base      & 75\% & 31.460 (+0.000)      \\
        \rowcolor{gray!30} +MaskBlur  & 75\% &  \textbf{31.616 \green{(+0.156)}}      \\
        \midrule
        InterNet-base      & 100\% & 31.596 (+0.000)      \\
        \rowcolor{gray!30} +MaskBlur  & 100\% & \textbf{31.732 \green{(+0.136)}}      \\
        \bottomrule
    \end{tabular}
    \label{tab:abl_size}
\end{table}

\begin{table}[tb]
    \centering
    
    \caption{
     Comprehensive comparison of  ATO \cite{jin2020light} and InterNet \cite{wang2020spatial} combined with different DA strategies in SR 4$\times$ tasks on five benchmark datasets. We report the average metric in terms of PSNR (dB, $\uparrow$) and SSIM ($\uparrow$), where $\delta$ indicates the performance gap from the baseline. The training time per epoch is also reported to represent the computation cost. Our results are marked in \colorbox {gray!30}{gray}.}
    \renewcommand\arraystretch{1.0}
    \resizebox{1 \columnwidth}{!}{
    \begin{tabular}{lccc}
        \toprule
        Method  & Training Time & Avg. PSNR ($\delta$) & Avg. SSIM ($\delta$) \\
        \midrule
        ATO-base      & 27 min & 31.494 (+0.000) & 0.9384 (+0.0000)      \\
        ATO-CutBlur   & 29 min & 31.576 \green{(+0.072)} & 0.9387 \green{(+0.0003)} \\
        ATO-CutMIB    & 29 min & 31.646 \green{(+0.142)} & 0.9390 \green{(+0.0006)} \\
        \rowcolor{gray!30} ATO-MaskBlur  & 29 min & 31.872 \green{(+0.378)} & 0.9401 \green{(+0.0017)}      \\
        \midrule
        InterNet-base      & 14 min & 31.596 (+0.000) & 0.9374 (+0.0000)      \\
        InterNet-CutBlur   & 15 min & 31.629 \green{(+0.033)} & 0.9381 \green{(+0.0007)} \\
        InterNet-CutMIB    & 15 min & 31.664 \green{(+0.068)} & 0.9389 \green{(+0.0015)} \\
        \rowcolor{gray!30} InterNet-MaskBlur  & 15 min & 31.732 \green{(+0.136)} & 0.9404 \green{(+0.0030)}       \\
        \bottomrule
    \end{tabular}
    }
    \label{tab:quanti_ssim}
\end{table}

\begin{table*}[tb]
\centering
\caption{Quantitative detailed comparison of various SR methods combined with different DA strategies in SR 4$\times$ tasks on benchmark datasets. We report the average PSNR (dB, $\uparrow$) and $\delta$ indicates the performance gap from the baseline.}
\resizebox{2 \columnwidth}{!}{
\begin{tabular}{lcccccc}
\toprule
  \multicolumn{1}{l}{\multirow{2.5}{*}{\textbf{Method} }}  &  \textbf{EPFL} & \textbf{HCInew} & \textbf{HCIold} & \textbf{INRIA} & \textbf{STFgantry} & \textbf{Average} \\ 
\cmidrule(lr){2-7}
  & PSNR ($\delta$) & PSNR ($\delta$) & PSNR ($\delta$) & PSNR ($\delta$) & PSNR ($\delta$)  & PSNR ($\delta$) \\
  
\midrule
 ATO-base & 28.515 (+0.000) & 30.813 (+0.000) & 36.893 (+0.000) & 30.677 (+0.000) & 30.573 (+0.000) & 31.494 (+0.000) \\
  ATO-CutBlur & 28.612 \green{(+0.097)} & 30.896 \green{(+0.083)} & 36.991 \green{(+0.098)} & 30.798 \green{(+0.121)} & 30.585 \green{(+0.012)} & 31.576 \green{(+0.072)} \\
 ATO-CutMIB & 28.631 \green{(+0.116)} & 30.950 \green{(+0.138)} & 37.050 \green{(+0.157)} & 30.829 \green{(+0.152)} & 30.771 \green{(+0.198)} & 31.646 \green{(+0.142)} \\
 \rowcolor{gray!30} ATO-MaskBlur & 28.893 \green{(+0.378)} & 31.075 \green{(+0.262)} & 37.274 \green{(+0.381)} & 31.089 \green{(+0.412)}  & 31.027 \green{(+0.454)} & 31.872 \green{(+0.378)} \\
 \midrule
 InterNet-base & 28.687 (+0.000) & 31.011 (+0.000) & 37.109 (+0.000) & 30.658 (+0.000) & 30.514 (+0.000) & 31.596 (+0.000) \\
 InterNet-CutBlur & 28.845 \green{(+0.158)} & 30.980 \red{(-0.031)} & 37.149 \green{(+0.040)} & 30.800 \green{(+0.142)} & 30.371 \red{(-0.143)} & 31.629 \green{(+0.033)} \\
 InterNet-CutMIB & 28.856 \green{(+0.169)} & 31.009 \red{(-0.002)} & 37.184 \green{(+0.075)} & 30.813 \green{(+0.155)} & 30.460 \red{(-0.054)} & 31.664 \green{(+0.068)} \\
 \rowcolor{gray!30} InterNet-MaskBlur & 28.811 \green{(+0.124)} & 31.108 \green{(+0.097)} & 37.331 \green{(+0.222)} & 30.733 \green{(+0.075)} & 30.679 \green{(+0.165)} & 31.732 \green{(+0.136)} \\
 \midrule
 IINet-base & 29.005 (+0.000) & 31.313 (+0.000) & 37.547 (+0.000) & 31.086 (+0.000) & 31.198 (+0.000) & 32.030 (+0.000) \\
 IINet-CutBlur & 29.046 \green{(+0.041)} & 31.357 \green{(+0.044)} & 37.595 \green{(+0.048)} & 31.026 \red{(-0.060)} & 31.300 \green{(+0.102)} & 32.065 \green{(+0.035)} \\
 IINet-CutMIB & 29.106 \green{(+0.101)} & 31.422 \green{(+0.109)} & 37.613 \green{(+0.066)} & 31.089 \green{(+0.003)} & 31.450 \green{(+0.252)} & 32.136 \green{(+0.106)} \\
 \rowcolor{gray!30} IINet-MaskBlur & 29.115 \green{(+0.110)} & 31.363 \green{(+0.050)} & 37.711 \green{(+0.164)} & 31.116 \green{(+0.030)} & 31.395 \green{(+0.197)} & 32.140 \green{(+0.110)}  \\
 \midrule
 DistgSSR-base & 29.015 (+0.000) & 31.410 (+0.000) & 37.588 (+0.000) & 31.015 (+0.000) & 31.635 (+0.000) & 32.133 (+0.000) \\
 DistgSSR-CutBlur &  29.023 \green{(+0.008)} & 31.418 \green{(+0.008)} & 37.599 \green{(+0.011)} & 31.022 \green{(+0.007)} & 31.638 \green{(+0.003)} & 32.140 \green{(+0.007)} \\
 DistgSSR-CutMIB & 29.034 \green{(+0.019)} & 31.445 \green{(+0.035)} & 37.618 \green{(+0.030)} & 31.030 \green{(+0.015)} & 31.681 \green{(+0.046)} & 32.162 \green{(+0.029)} \\
 \rowcolor{gray!30} DistgSSR-MaskBlur & 29.282 \green{(+0.267)} & 31.453 \green{(+0.043)} & 37.656 \green{(+0.068)} & 31.304 \green{(+0.289)} & 31.702 \green{(+0.067)} & 32.279 \green{(+0.146)} \\
 \midrule
 EPIT-base & 29.317 (+0.000) & 31.510 (+0.000) & 37.679 (+0.000) & 31.352 (+0.000) & 32.180 (+0.000) & 32.408 (+0.000) \\
  EPIT-CutBlur & 29.288 \red{(-0.029)} & 31.504 \red{(-0.006)} & 37.677 \red{(-0.002)} & 31.287 \red{(-0.065)} & 32.175 \red{(-0.005)} & 32.386 \red{(-0.022)} \\ % need updated by cwt
 EPIT-CutMIB &  29.309 \red{(-0.008)} & 31.514 \green{(+0.004)} & 37.681 \green{(+0.002)} & 31.332 \red{(-0.020)} & 32.192 \green{(+0.012)} & 32.406 \red{(-0.002)} \\
 \rowcolor{gray!30} EPIT-MaskBlur & 29.328 \green{(+0.011)}  & 31.528 \green{(+0.018)} & 37.703 \green{(+0.024)} & 31.384 \green{(+0.032)}  & 32.202 \green{(+0.004)} & 32.429 \green{(+0.021)} \\ 
\bottomrule
\end{tabular}
}

\label{table:quantitative}
\end{table*}

% \subsubsection{Spatial mask block-wise size.}
% \label{sec: spa_patch}
\noindent \textbf{Spatial mask block-wise size.}  We investigate the impact of the block-wise size of the spatial mask while using training image patches of size 32$\times$32. As evident from Table \ref{tab:abl_spa}, a minimal block-wise size can disrupt the local completeness of information, while a tremendous block-wise size results in increased information loss, e.g., the information all comes from the LR domain, making it difficult for the model to learn. Our experiments determined that a block-wise size of 4 is optimal, as it strikes a balance between local completeness and information loss.

% \subsubsection{Spatial mask ratios.}
% \label{sec: spa_ratio}
\noindent \textbf{Spatial mask ratio.} We investigate the impact of the spatial mask ratio. Table \ref{tab:abl_spa} presents the results for different spatial masking ratios, with a ratio of 50\% yielding the best performance. Our analysis suggests that when the masking ratio is either too high or too low, the new training samples contain only a small portion of either LR or HR information, leading to suboptimal spatial DA. Therefore, a 50\% ratio provides a balanced mix of LR and HR information in the new training samples, resulting in the most effective spatial DA.

% \subsubsection{Angular mask strategies.}
% \label{sec: ang_strategy}
\noindent \textbf{Angular mask sampling strategy.} We analyze some factors that affect angular dropout. Similar to spatial mask sampling strategies, we first investigate different angular mask sampling strategies, i.e., grid sampling and random sampling, as shown in Fig.~\ref{fig:ang_mask}. From Table~\ref{tab:abl_ang}, we observe that both sampling strategies can further improve performance based on spatial blur, with random sampling achieving greater improvement. Furthermore, we design special angular masks composed of different angular directions and their combinations, i.e., horizontal, vertical, main diagonal, and antidiagonal, as shown in Fig.~\ref{fig:ang_mask}. Table~\ref{tab:abl_ang} demonstrates that the effectiveness of special angular masks is significantly lower compared to random sampling, which indicates that special angular masks have little impact on angular dropout.

% \subsubsection{Angular mask sampling ratios.}
% \label{sec: ang_ratio}
\noindent \textbf{Angular mask ratio.} 
The angular masking ratio also affects the performance of angular dropout, so we explore various ratios to understand their impact. As shown in Table~\ref{tab:abl_ang}, the optimal performance is achieved when the ratio is set to 75\%, indicating that when 75\% of the views are dropped out while the remaining 25\% undergo spatial blur, the performance is optimal. The finding differs from the spatial domain, suggesting that retaining a higher proportion of LR information during the DA process is crucial in the angular domain.

% \subsubsection{Angular dropout combine with other DA strategies.}
% \label{sec: other}
\noindent \textbf{Angular dropout combined with other spatial DA strategies.} Our MaskBlur consists of two separate operations: spatial blur and angular dropout. Therefore, angular dropout can also be combined with other spatial DA strategies. We compare several general spatial DA strategies combined with angular dropout, i.e., RGB permute, Cutout~\cite{devries2017improved}, Mixup~\cite{zhang2017mixup}, and CutBlur~\cite{yoo2020rethinking}. From Table ~\ref{tab:abl_comb}, we can find that while the general spatial DA strategies can improve the performance of the baseline model, their impact is comparatively modest and falls short of the improvement achieved by spatial blur alone. Moreover, when these spatial DA strategies are combined with angular dropout, the improvement becomes notably more substantial. This result underscores the broad applicability and generalization of our angular dropout.

\noindent \textbf{DA probability.} DA probability $p$ determines the ratio of samples trained with MaskBlur. We conduct experiments to choose DA probability $p$ under the condition that the spatial mask and angular mask adopt the random sampling strategy, spatial mask block-wise size is set to 4, the spatial masking ratio is set to 0.5, and the angular masking ratio is set to 0.75. Table~\ref{tab:abl_prob} shows that the model can achieve the highest PSNR at $p=0.25$. We analyze some differences between the augmented sample distribution and the original sample distribution. Therefore, a high proportion of DA will destroy the original image distribution, resulting in sub-optimal results, while using a low proportion of DA can simultaneously maintain the original sample distribution and increase the training sample, resulting in better results.

\begin{figure*}[!tb]
  \centering
  \includegraphics[width=\linewidth]{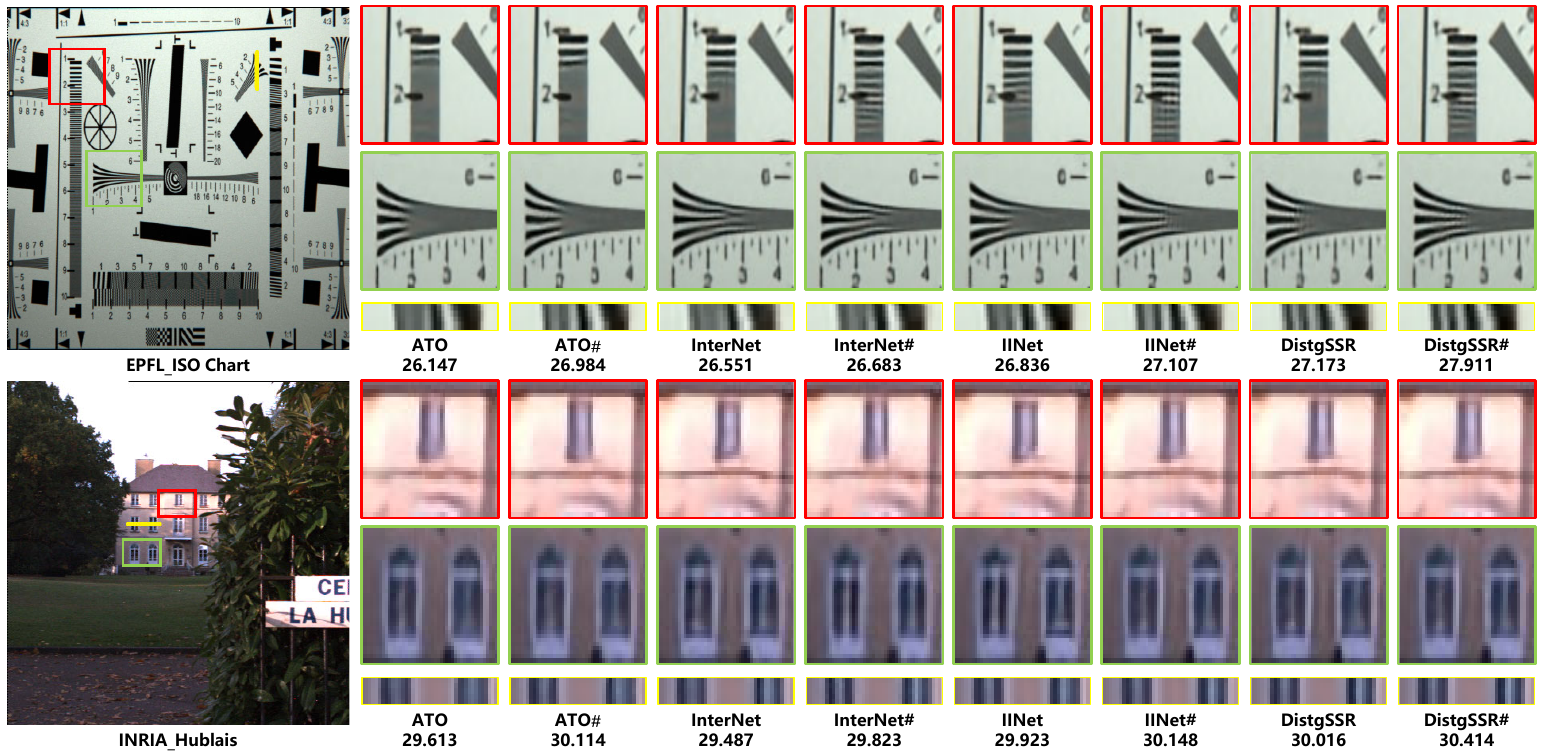}
  \caption{Qualitative comparisons of various SR models trained with and without the MaskBlur 4$\times$ LF image SR. $\#$ denotes that the networks are trained with the MaskBlur. We report the PSNR (dB, $\uparrow$). Please zoom in for better visualization.}
  \label{fig:sr_compare}
\end{figure*}

\begin{figure}[tb]
  \centering
  \includegraphics[width=\linewidth]{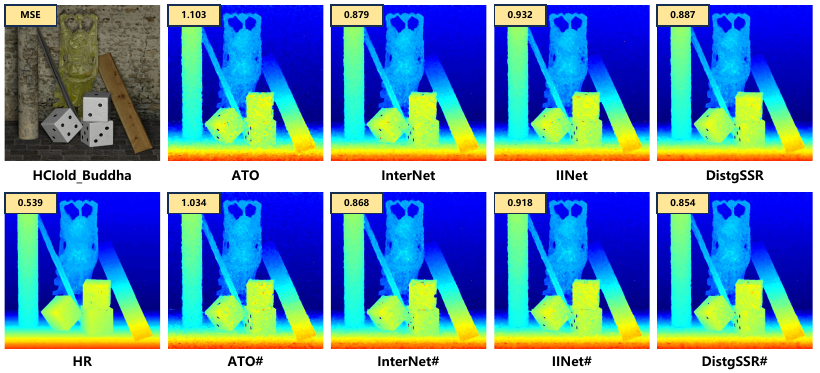}
  \caption{Depth estimation results achieved by SPO~\cite{zhang2016robust} using 4$\times$ SR LF images produced by different SR methods trained with and without the MaskBlur. $\#$ denotes that the networks are trained with the MaskBlur. We report the mean square error multiplied by 100 (MSE $\times$100), lower is better.}
  \label{fig:depth_compare}
\end{figure}

\begin{table}[!tb]
\centering
% \caption{Quantitative comparison of various methods for LF image denoising task with different noise levels. We report the average PSNR (dB, $\uparrow$) and $\delta$ indicates the performance gap from the baseline.}
\caption{Quantitative comparison of various methods for LF image denoising task with different noise levels on HCInew \cite{honauer2016dataset}. We report the average PSNR (dB, $\uparrow$) and $\delta$ indicates the performance gap from the baseline. The models trained with MakBlur are marked in \colorbox {gray! 30}{gray}.}
\renewcommand\arraystretch{1.0}
\resizebox{1 \columnwidth}{!}{
\begin{tabular}{lccc}
\toprule
\multicolumn{1}{l}{\multirow{2.5}{*}{\textbf{Method}}}  &  \textbf{$\sigma=10$} & \textbf{$\sigma=20$} & \textbf{$\sigma=50$} \\ 
\cmidrule(lr){2-4}
 & PSNR ($\delta$)  & PSNR ($\delta$)  & PSNR ($\delta$)  \\
\midrule
ATO & 38.422 (+0.000) & 35.152 (+0.000) & 31.343 (+0.000) \\
\rowcolor{gray!30}  +MaskBlur & 38.546 \green{(+0.124)} & 35.281 \green{(+0.129)} & 31.535 \green{(+0.192)}  \\
\midrule
InterNet & 38.260 (+0.000) & 35.493 (+0.000) & 31.980 (+0.000) \\
\rowcolor{gray!30}  +MaskBlur & 38.560 \green{(+0.300)} & 35.548 \green{(+0.055)} & 32.090 \green{(+0.110)}  \\
\midrule
IINet & 39.526 (+0.000) & 36.466 (+0.000) & 32.258 (+0.000) \\
\rowcolor{gray!30}  +MaskBlur & 39.566 \green{(+0.040)} & 36.512 \green{(+0.046)} & 32.324 \green{(+0.066)}  \\
\midrule
DistgSSR & 39.662 (+0.000) & 36.617 (+0.000) &  32.449 (+0.000)\\
\rowcolor{gray!30}  +MaskBlur &  39.785 \green{(+0.123)} & 36.630 \green{(+0.013)} & 32.496 \green{(+0.047)}  \\
\bottomrule
\end{tabular}
}

\label{table:dn_compare}
\end{table}

\begin{table}[!t]
\centering
\caption{Quantitative comparison of various methods for LF image deblurring task with different blur kernels on HCInew \cite{honauer2016dataset}. We report the average PSNR (dB, $\uparrow$) and $\delta$ indicates the performance gap from the baseline. The models trained with MakBlur are marked in \colorbox {gray! 30}{gray}.}
% \caption{Quantitative comparison of various methods for LF image deblurring task with different blur kernels.}
\renewcommand\arraystretch{1.0}
\resizebox{1 \columnwidth}{!}{
\begin{tabular}{lccc}
\toprule
% \multicolumn{1}{l}{\multirow{2}{*}{}}
% \multirow{1}{*}{\diagbox{\textbf{Method}}{\textbf{Blur}}} 
% \diagbox{\textbf{Method}}{\textbf{Blur}} & 
% \begin{minipage}[b]{0.15\columnwidth}
%     \centering
%     \raisebox{-.5\height}{\includegraphics[width=\linewidth]{figs/kernel_02.jpg}}
% \end{minipage}
% &
% \begin{minipage}[b]{0.15\columnwidth}
%     \centering
%     \raisebox{-.5\height}{\includegraphics[width=\linewidth]{figs/kernel_04.jpg}}
% \end{minipage}
% &
% \begin{minipage}[b]{0.15\columnwidth}
%     \centering
%     \raisebox{-.5\height}{\includegraphics[width=\linewidth]{figs/kernel_08.jpg}}
% \end{minipage}
% \\ 
\multicolumn{1}{l}{\multirow{2.5}{*}{\textbf{Method}}}  &  \textbf{isotropic} & \textbf{anisotropic} & \textbf{motion} \\ 
\cmidrule{2-4}
 & PSNR ($\delta$)  & PSNR ($\delta$)  & PSNR ($\delta$)  \\
\midrule
ATO & 30.018 (+0.000) & 30.947 (+0.000) & 23.450 (+0.000) \\
\rowcolor{gray!30}  +MaskBlur & 30.170 \green{(+0.152)} & 31.027 \green{(+0.080)} & 23.651 \green{(+0.201)}  \\
\midrule
InterNet & 30.709 (+0.000) & 29.678 (+0.000) & 23.040 (+0.000) \\
\rowcolor{gray!30}  +MaskBlur & 31.027 \green{(+0.318)} & 29.872 \green{(+0.194)} & 23.127 \green{(+0.087)} \\
\midrule
IINet & 31.652 (+0.000) & 30.849 (+0.000) & 23.335 (+0.000) \\
\rowcolor{gray!30}  +MaskBlur & 32.009 \green{(+0.357)} & 31.119 \green{(+0.270)} & 23.455 \green{(+0.120)} \\
\midrule
DistgSSR & 32.776 (+0.000) & 31.287 (+0.000) & 24.131 (+0.000) \\ % need update by cwt
\rowcolor{gray!30}  +MaskBlur &  32.908 \green{(+0.132)} &  31.332 \green{(+0.045)} &  24.656 \green{(+0.039)}  \\
\bottomrule
\end{tabular}
}

\label{table:db_compare}
\end{table}

\noindent \textbf{Various dataset size.} We further investigate the model performance under different dataset sizes, as shown in Table~\ref{tab:abl_size}. We conduct experiments with 100\%, 75\%, 50\%, and 25\% of the dataset size, respectively. First, our method brings great improvements to the Internet in various settings. As the ratio of the dataset decreases, the performance gap between the baseline and our method widens. When the dataset ratio is less than 50\%, our method demonstrates an improvement of up to 0.201 dB. In addition, when trained with MaskBlur, the InterNet trained on 75\% of the dataset outperforms the 100\% baseline. This proves the effectiveness of our method. 
% Our method also significantly alleviates the overfitting problem (see Figure 1). For example, if we use 25\% of the training data, the large model, i.e., InterNet is prone to overfitting, which can be significantly reduced by using our method. 

\begin{figure}[tb]
  \centering
  \includegraphics[width=\linewidth]{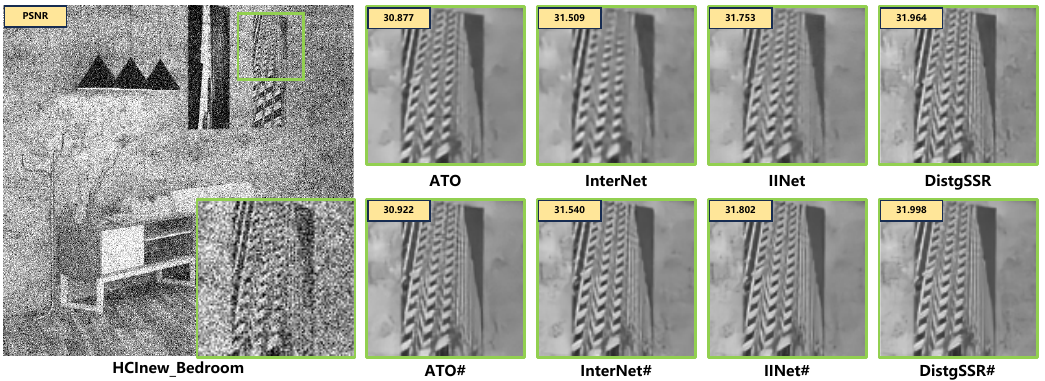}
  \caption{Visual comparisons of various models trained with and without MaskBlur on the $\sigma = 50$ LF denoising task. We report the PSNR (dB, $\uparrow$) on the top-left.}
  \label{fig:dn_compare}
\end{figure}

\begin{figure}[tb]
  \centering
  \includegraphics[width=\linewidth]{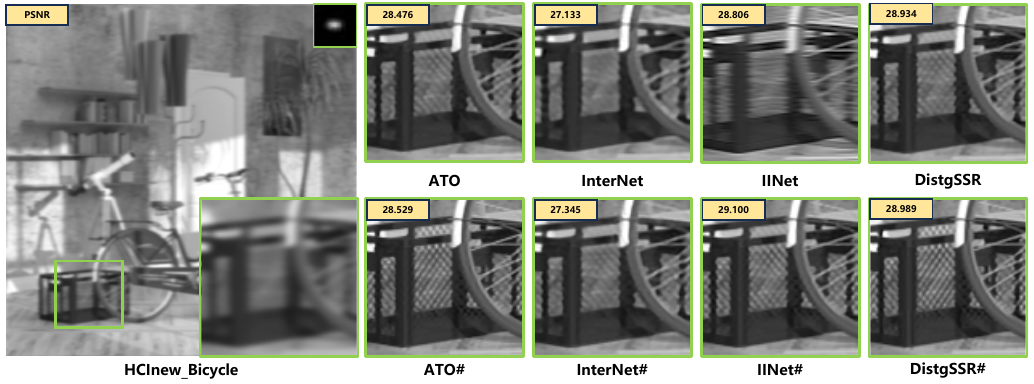}
  \caption{Visual comparisons of various models with and without MaskBlur on LF deblurring under the anisotropic Gaussian blur. We report the PSNR (dB, $\uparrow$) on the top-left.}
  \label{fig:db_compare}
\end{figure}

\begin{figure}[t]
  \centering
  \includegraphics[width=\linewidth]{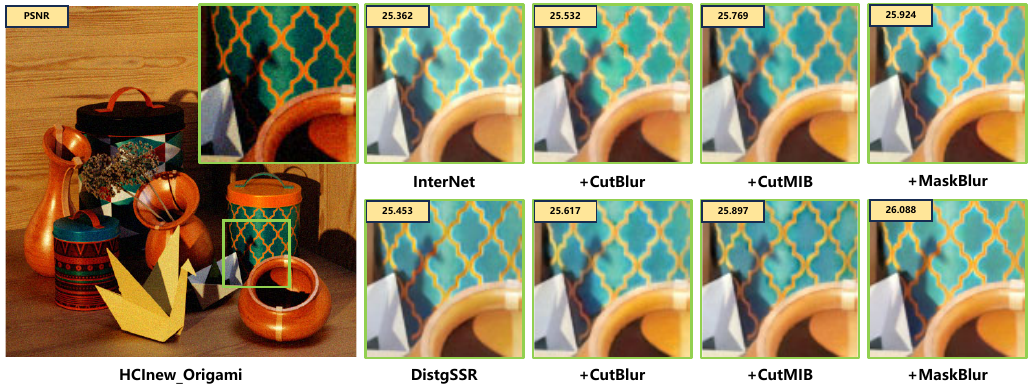}
  \caption{Visual comparisons of various models trained with different DA schemes on the $\gamma = 0.3$ LF low-light enhancement task. We report the PSNR (dB, $\uparrow$) on the top-left.}
  \label{fig:llie_compare}
\end{figure}

\begin{figure}[tb]
  \centering
  \includegraphics[width=0.95 \linewidth]{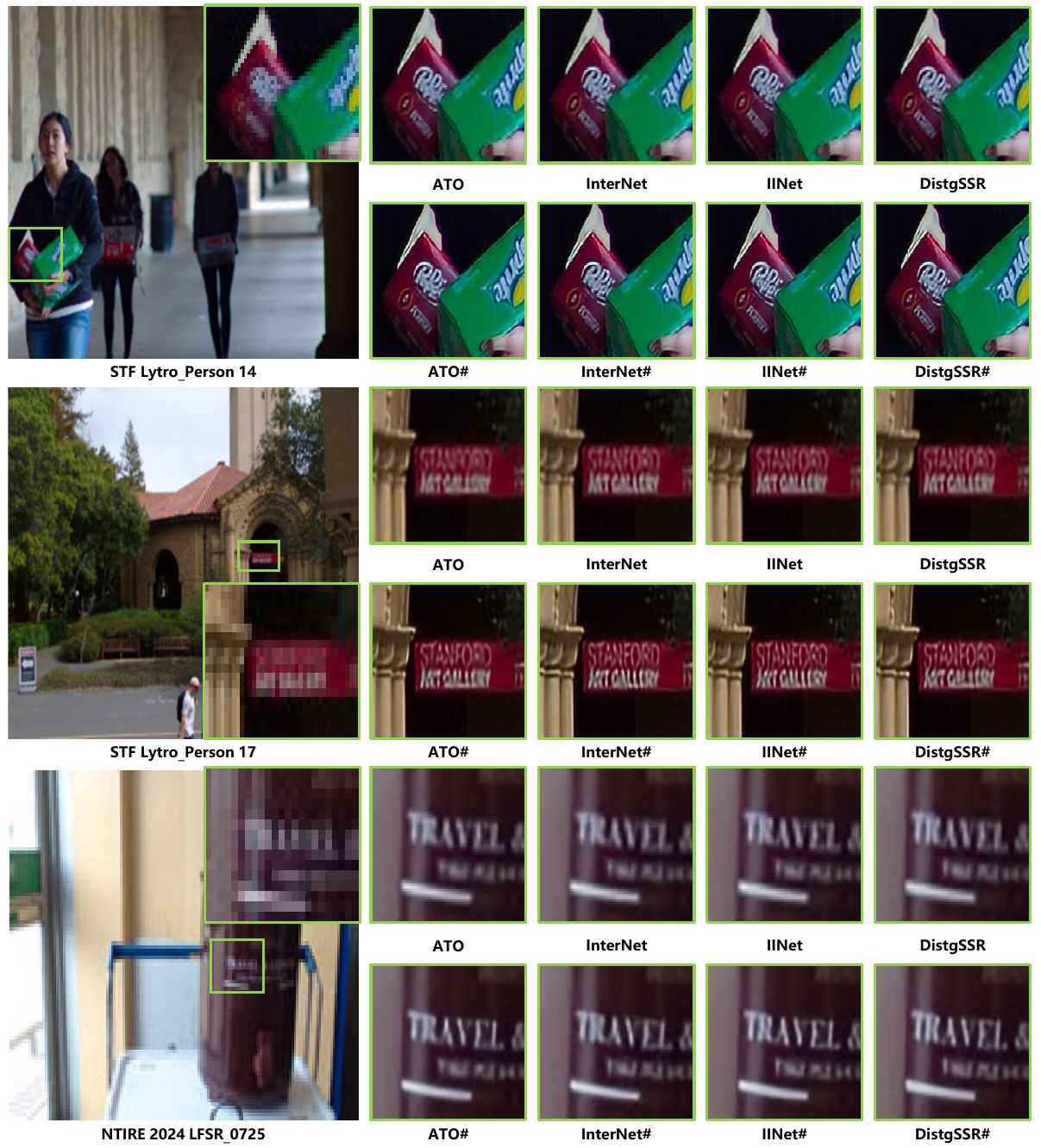}
  % \includegraphics[width=\linewidth]{figs/fig_noise_compare3.pdf}
  
  % \caption{Visual comparisons of various models with and without MaskBlur on a real-world scene \textit{Person 14} from STF Lytro \protect \cite{SLFA}.}
  \caption{Visual comparisons of various models with and without MaskBlur on real-world scenes from STF Lytro~\cite{SLFA} and NTIRE 2024 LFSR validation set~\cite{NTIRE2024-LFSR}}.
  \label{fig:real_compare}
\end{figure}

\subsection{Quantitative Comparisons}
\label{sec: quantita}

We compare MaskBlur with state-of-the-art (SOTA) DA strategies for SR, i.e., CutBlur~\cite{yoo2020rethinking} and CutMIB~\cite{xiao2023cutmib}, on various benchmark datasets. The results are shown in Tables~\ref{table:quantitative} and \ref{tab:quanti_ssim}. Both CutBlur and CutMIB strategies improve the average PSNR of the baseline methods, except for a slight performance decrease on a few individual datasets. We further analyzed the reason for the poor performance of EPIT-CutBlur where CutBlur applies data augmentation solely in the spatial domain without considering the angular domain. CutBlur degrades the quality of the EPI subspace significantly, thereby increasing the difficulty for EPIT to learn the spatial-angular correlations within the EPI subspace, leading to sub-optimal performance. 
As observed in Tables~\ref{table:quantitative} and \ref{tab:quanti_ssim}, compared to MaskBlur, 
CutBlur and CutMIB have limited improvements to the models. For instance, the ATO model trained with MaskBlur achieves an average PSNR improvement of \textbf{0.378 dB} compared to the baseline. The improvement with CutBlur and CutMIB is only 0.072 dB and 0.142 dB, respectively. The InterNet trained with MaskBlur achieves an average gain of \textbf{0.0030} in terms of SSIM than CutBlur (0.0007) and CutMIB (0.0015).
We also compared the computational costs with other DA schemes, as shown in Table ~\ref{tab:quanti_ssim}. 
While MaskBlur is comparable to CutBlur and CutMIB in terms of computational costs, it significantly outperforms them in terms of performance. The computational overhead introduced by MaskBlur during training is minimal. For instance, when applied to the InterNet model, the training time per epoch increases by approximately 1 minute. This slight increase in training time is worthwhile considering the performance improvements MaskBlur provides. Importantly, MaskBlur is only executed during the training phase. Therefore, it does not add any computational overhead during the inference phase. When combined with other SR methods, MaskBlur demonstrates superior overall performance compared to CutBlur and CutMIB, confirming the effectiveness of our approach.

\begin{table}[t]
    \centering
    \caption{Quantitative comparison of various methods trained with different DA schemes in LF image low-light enhancement task under different gamma correction on HCInew \cite{honauer2016dataset}. We report the average PSNR (dB, $\uparrow$) and $\delta$ indicates the performance gap from the baseline. The models trained with MakBlur are marked in \colorbox {gray! 30}{gray}.}
    \renewcommand\arraystretch{1.0}
    \resizebox{1 \columnwidth}{!}{
    \begin{tabular}{lccc}
    \toprule
    \multicolumn{1}{l}{\multirow{2.5}{*}{\textbf{Method}}}  &  \textbf{$\gamma=0.2$} & \textbf{$\gamma=0.3$} & \textbf{$\gamma=0.5$} \\ 
    \cmidrule(lr){2-4}
     & PSNR ($\delta$)  & PSNR ($\delta$)  & PSNR ($\delta$)  \\
    \midrule
    InterNet & 24.222 (+0.000) & 28.898 (+0.000) & 33.161 (+0.000) \\
    +CutBlur &  24.284 \green{(+0.062)} & 28.942 \green{(+0.044)} & 33.185 \green{(+0.024)} \\
    +CutMIB &  24.353 \green{(+0.131)} & 29.014 \green{(+0.116)} & 33.211 \green{+0.050} \\
    \rowcolor{gray!30}  +MaskBlur & 24.575 \green{(+0.353)} & 29.121 \green{(+0.223)} & 33.285 \green{(+0.124)}  \\
    \midrule
    DistgSSR & 25.526 (+0.000) &  29.480 (+0.000) & 34.804 (+0.000) \\
    +CutBlur & 25.530 \green{(+0.004)}  & 29.492 \green{(+0.012)} & 34.814 \green{(+0.010)} \\
    +CutMIB &  25.632 \green{(+0.106)} & 29.598 \green{(+0.118)} & 34.920 \green{(+0.116)}\\
    \rowcolor{gray!30}  +MaskBlur & 25.821 \green{(+0.295)} & 29.673 \green{(+0.193)} & 34.959 \green{(+0.155)}  \\
    \bottomrule
    \end{tabular}
    }
\label{table:llie_compare}
\end{table}

\begin{table}[tb]
    \centering
    \caption{Quantitative comparison of various methods in real-world LF image 4$\times$ SR task under different isotropic Gaussian kernels on EPFL \cite{rerabek2016new}. We report the average PSNR (dB, $\uparrow$) and $\delta$ indicates the performance gap from the baseline. The models trained with MakBlur are marked in \colorbox {gray! 30}{gray}.}
    \renewcommand\arraystretch{1.0}
    \resizebox{1 \columnwidth}{!}{
    \begin{tabular}{lccc}
    \toprule
    \multicolumn{1}{l}{\multirow{2.5}{*}{\textbf{Method}}}  &  \textbf{$k=1.8$} & \textbf{$k=2.5$} & \textbf{$k=3.2$} \\ 
    \cmidrule(lr){2-4}
     & PSNR ($\delta$)  & PSNR ($\delta$)  & PSNR ($\delta$)  \\
    \midrule
    				
    ATO & 25.084 (+0.000) & 23.630 (+0.000) & 22.638 (+0.000) \\
    \rowcolor{gray!30}  +MaskBlur & 25.096 \green{(+0.012)} & 23.651 \green{(+0.021)} & 22.658 \green{(+0.020)}  \\
    \midrule
    InterNet & 25.168 (+0.000) & 23.646 (+0.000) & 22.638 (+0.000) \\
    \rowcolor{gray!30}  +MaskBlur & 25.149 \green{(+0.002)} & 23.666 \green{(+0.020)} & 22.656 \green{(+0.018)}  \\
    \midrule
    IINet & 25.165 (+0.000) &  23.650 (+0.000) & 22.634 (+0.000) \\
    \rowcolor{gray!30}  +MaskBlur & 25.171 \green{(+0.006)} & 23.652 \green{(+0.002)} & 22.645 \green{(+0.011)}  \\
    \midrule
    DistgSSR & 25.111 (+0.000) &  23.650 (+0.000) & 22.641 (+0.000) \\
    \rowcolor{gray!30}  +MaskBlur & 25.151 \green{(+0.040)} & 23.665 \green{(+0.015)} & 22.651 \green{(+0.010)}  \\
    \bottomrule
    \end{tabular}
    }

\label{table:blind_compare}
\end{table}

\subsection{Qualitative Comparisons}
\label{sec: qualita}

% \textbf{SR image visualization comparisons.}

% \textbf{Depth map visualization comparison.}

We also qualitatively compare the results of various baseline models with and without MaskBlur on 4$\times$ LF image SR task, as shown in Fig.~\ref{fig:sr_compare}. Firstly, we can observe a noticeable improvement in spatial visual quality when using the MaskBlur DA strategy, e.g., the ruling lines appear clearer. Secondly, the EPI images with MaskBlur exhibit clear pixel slopes, reflecting the ability of our method to maintain angular consistency. 
Furthermore, following previous methods~\cite{wang2022disentangling,xiao2023cutmib}, we apply a depth estimation algorithm, i.e., SPO~\cite{zhang2016robust} to the super-resolved LF images, and the results are shown in Fig.~\ref{fig:depth_compare}. The depth maps obtained employing the MaskBlur strategy show smaller average errors (MSE $\times$100) and provide more accurate estimations in some edge regions. This further demonstrates that our MaskBlur can maintain angular consistency.

\subsection{Applications}
\label{sec: apply}

\noindent \textbf{LF image denoising.} We extend MaskBlur to the LF image denoising task. Following~\cite{chen2018light,guo2021deep}, we utilize the same noise generation strategy to create the HCInew \cite{honauer2016dataset} dataset for training and testing. Specifically, we add zero-mean Gaussian noise with different standard deviations ($\sigma=10, 20, 50$) to the LF images to generate the noisy images. Next, we remove the upsampling operation from the SR models and retrain the ATO, InterNet, IINet, and DistgSSR models for different noise levels. Table~\ref{table:dn_compare} presents the quantitative results with and without the MaskBlur DA strategy. We can observe that MaskBlur consistently improves the PSNR performance of the baseline models across different noise settings. As shown in Fig.~\ref{fig:dn_compare}, the denoised images with MaskBlur exhibit less noise and clearer lines than those without MaskBlur.

\noindent \textbf{LF image deblurring.} We evaluate the effectiveness of MaskBlur on the LF image deblurring task. Specifically, we synthesize blurred images using isotropic Gaussian kernel, anisotropic Gaussian blur, and motion blur. Similar to LF image denoising, we retrain the ATO, InterNet, IINet, and DistgSSR models for different blur types on the HCInew \cite{honauer2016dataset} dataset. Table~\ref{table:db_compare} presents the quantitative results for the three different types of blur kernels. We can observe that the models trained with the MaskBlur strategy achieve better PSNR results. We also visualize the deblurring results in Figure~\ref{fig:db_compare}, and it can be seen that the deblurred images generated by the models trained with MaskBlur effectively reduce blur and improve the visual quality of the views.

\noindent \textbf{LF image low-light enhancement.} We also compare MaskBlur with other DA schemes, i.e., CutBlur~\cite{yoo2020rethinking} and CutMIB~\cite{xiao2023cutmib}, for the task of LF image low-light enhancement. Specifically, inspired by~\cite{park2017low,lore2017llnet,zhang2023lrt}, we synthesize low-light LF images using different level gamma correction ($\gamma=0.2, 0.3, 0.5$) and add zero-mean Gaussian noise with deviations ($\sigma=10$). We conduct experiments on the HCInew~\cite{honauer2016dataset} dataset using the InterNet~\cite{wang2020spatial} and DistgSSR~\cite{wang2022disentangling}. Table \ref{table:llie_compare} shows the quantitative comparison results, demonstrating that training InterNet and DistgSSR with MaskBlur performs better than other DA schemes. Additionally, as shown in Fig.~\ref{fig:llie_compare}, models trained with the MaskBlur produce clearer edges and textures, further proving the generalization capability of the MaskBlur scheme.

% \subsection{Blind LF image SR}
\noindent \textbf{Real-world LF image SR.} Most existing LF image SR methods employ bicubic interpolation to simulate the degradation between HR and LR. However, the degradation of real-world LF images is very complicated. The mathematical equation of the classical degradation model~\cite{liu2013bayesian,wang2024real} is given below.

\begin{equation}
\mathcal{I} _{i}^{LR}= (\mathcal{I} _{i}^{HR}\otimes k_i)\downarrow _s+n_i
\end{equation}

\noindent where $\otimes$ represents convolution operation, $k_i$ is a Gaussian blur kernel, $\downarrow _s$ is the downsampling operation with scale factor $s$, and $n_i$ represents Gaussian noise.
Bicubic degradation can be considered a special case of classical degradation as it can be approximated by setting an appropriate zero-noise kernel\cite{bell2019blind,zhang2020deep}. Various factors, such as blur kernel and noise level influence the degradation model. 

We employ the classical degradation model to simulate real scene degradation and evaluate the generalization of our method in real-world LF image SR task on EPFL~\cite{rerabek2016new} since the dataset contains real-world scenes captured by a Lytro Illum LF camera.
Specifically, we directly utilize the baseline networks, i.e., InterNet \cite{wang2020spatial}, ATO \cite{jin2020light}, IINet \cite{liu2021intra}, DistgSSR \cite{wang2022disentangling}, and their corresponding versions trained with MaskBlur to conduct real-world SR. 
Note that we only consider isotropic Gaussian blur kernel, with each view being assigned the same degenerate kernel, and noise is not considered. Following the setting in~\cite{gu2019blind,huang2020unfolding}, the kernel size is set to 21 and the kernel width is set to 1.8, 2.5, and 3.2 for evaluation.
As shown in Table~\ref{table:blind_compare}, various models trained with MaskBlur consistently exhibit performance improvements across different Gaussian blur kernels. This underscores the generalization ability of our method in the real-world LF image SR task.

Furthermore, we compare different methods trained with and without the MaskBlur on real-world LF images taken with a Lytro Illum camera from STF Lytro~\cite{SLFA} and NTIRE 2024 LFSR validation set~\cite{NTIRE2024-LFSR}. Since there is no ground truth available for real-world scenes, we perform a visual comparison. Figure~\ref{fig:real_compare} shows that the models with the MaskBlur strategy generate better detail preservation and visual effects, which demonstrates the great generalization ability of our MaskBlur.

\section{Conclusion}
\label{sec: conclu}
This paper has proposed a novel DA strategy, MaskBlur, for LF image SR, utilizing both spatial and angular masks. The spatial mask dictates where pixels undergo blurring, while the angular mask determines which views are dropped. MaskBlur encourages the model to adaptively treat pixels differently in the spatial and angular domains during LF image SR. The effectiveness of MaskBlur is demonstrated through notable performance improvements in existing SR models. Furthermore, we validate the generalization of MaskBlur in LF image denoising, deblurring, low-light enhancement, and real-world SR tasks. Looking ahead, our future work will explore the application of MaskBlur for video-related tasks.

% \section{References Section}
% You can use a bibliography generated by BibTeX as a .bbl file.
%  BibTeX documentation can be easily obtained at:
%  http://mirror.ctan.org/biblio/bibtex/contrib/doc/
%  The IEEEtran BibTeX style support page is:
%  http://www.michaelshell.org/tex/ieeetran/bibtex/
 
%  % argument is your BibTeX string definitions and bibliography database(s)
% %\bibliography{IEEEabrv,../bib/paper}
% You can manually copy in the resultant .bbl file and set second argument of $\backslash${\tt{begin}} to the number of references
%  (used to reserve space for the reference number labels box).

\bibliographystyle{IEEEtran}
\bibliography{reference}

\end{document}